\documentclass[12pt]{article}
 \usepackage{epsfig}
 \def\be{\begin{equation}}
 \def\ee{\end{equation}}
 \def\bea{\begin{eqnarray}}
 \def\eea{\end{eqnarray}}
 \usepackage{graphicx}

 \catcode`\@=11
 \def\lsim{\mathrel{\mathpalette\@versim<}}
 \def\gsim{\mathrel{\mathpalette\@versim>}}
 \def\@versim#1#2{\vcenter{\offinterlineskip
 \ialign{$\m@th#1\hfil##\hfil$\crcr#2\crcr\sim\crcr } }}
 \catcode`\@=12

 \catcode`@=12
\begin{document}
 \thispagestyle{empty}
 \begin{flushright}
 UCRHEP-T588\\
 Mar 2018\
 \end{flushright}
 \vspace{0.6in}
 \begin{center}
 {\LARGE \bf Predestined Dark Matter in Gauge\\ 
Extensions of the Standard Model\\}
 \vspace{1.2in}
 {\bf Ernest Ma\\}
 \vspace{0.2in}
{\sl Physics and Astronomy Department,\\ 
University of California, Riverside, California 92521, USA\\}
\vspace{0.1in}
{\sl Jockey Club Institute for Advanced Study,\\ 
Hong Kong University of Science and Technology, Hong Kong, China\\} 
\end{center}
 \vspace{1.2in}

\begin{abstract}\
In any gauge extension of the standard model (SM) of quarks and leptons, 
there is a minimal set of fermion and scalar multiplets which encompasses 
all the particles and interactions of the SM.  Included within this set, 
there may be a suitable dark-matter candidate.  If not, one may still exist 
from the judicious addition of a simple fermion or scalar multiplet 
\underline{without any imposed symmetry}.  Some new examples of such 
predestined dark matter are discussed.
\end{abstract}

 \newpage
 \baselineskip 24pt

\noindent \underline{\it Introduction}~:~
Automatic (or predestined) symmetries play an important part in the structure 
of particle physics.  In the context of the standard 
$SU(3)_C \times SU(2)_L \times U(1)_Y$ gauge model of quarks and leptons (SM), 
baryon number $B$ and lepton number $L$ are automatically conserved 
because of the chosen particle content under the gauge symmetry. 
Furthermore, $CP$ conservation in the hadronic sector would also be 
automatic if there were only two families of quarks.  If the seesaw 
mechanism is invoked to obtain very small Majorana neutrino masses by 
increasing the particle content with the addition of heavy right-handed 
Majorana singlet neutrinos, then lepton number $L$ is replaced with 
lepton parity $(-1)^L$.  Suppose the particle content is increased again 
with a fermion or scalar multiplet, what new phenomena would occur? 

It was shown already some years ago~\cite{cfs06} that the neutral member of 
a fermion quintet or a scalar septet would be a good dark-matter candidate.  
Either addition to the SM \underline{without any imposed symmetry} would 
imply an automatic dark parity which keeps this neutral component (which 
is also automatically the lightest) against 
possible decay.  Whereas this is an attractive idea 
phenomenologically, the requirement of such large fundamnetal multiplets 
is not quite so theoretically.

In this paper, the notion of predestined dark matter is explored in the 
context of some left-right extensions of the SM, as well as 
gauge $U(1)$ extensions with heavy Majorana singlet or triplet fermions as the 
seesaw anchors of small neutrino masses.  It will be shown that simple 
solutions exist in both cases.  Some are already known and others are new. 

\noindent \underline{\it Minimal Left-Right Extension}~:~
Consider the well-known conventional left-right extension of the SM.  Under 
$SU(3)_C \times SU(2)_L \times SU(2)_R \times U(1)$,
\begin{eqnarray}
&&  \pmatrix{u \cr d}_L \sim (3,2,1,{1 \over 6}), ~~~ 
\pmatrix{u \cr d}_R \sim (3,1,2,{1 \over 6}), ~~~ \\ 
&&  \pmatrix{\nu \cr e}_L \sim (1,2,1,-{1 \over 2}), ~~~ 
\pmatrix{\nu \cr e}_R \sim (1,1,2,-{1 \over 2}).
\end{eqnarray}
The only fermion additions to the SM are the right-handed neutrinos $\nu_R$, 
and the $U(1)$ gauge symmetry is seen to be just $(B-L)/2$.  To obtain 
fermion masses, at least one scalar bidoublet 
\begin{equation}
\eta = \pmatrix{\eta_1^0 & \eta_2^+ \cr \eta_1^- & \eta_2^0} 
\sim (1,2,2,0)
\end{equation}
is needed. The choice of the rest of the scalar sector depends on how 
neutrinos acquire mass.  All possible such scenarios have been discussed 
some years ago~\cite{m04}.

Suppose an $SU(2)_R$ scalar doublet $\Phi_R = (\phi^+_R,\phi^0_R)$ is 
added~\cite{admnw09}, then the particle content of this model guarantees 
also $B$ and $L$ conservation, and neutrinos are Dirac particles. 
Suppose instead an $SU(2)_R$ scalar triplet 
$\xi_R = (\xi_R^{++}, \xi_R^+, \xi_R^0)$ is used, then $\nu_R$ acquires 
a large Majorana mass and $L$ is broken to $(-1)^L$, resulting in the 
canonical seesaw mechanism for small Majorana neutrino masses.

Consider now the addition of a fermion bidoublet 
\begin{equation}
\psi = \pmatrix{\psi^0_1 & \psi^+_2 \cr \psi^-_1 & \psi^0_2} \sim (1,2,2,0)
\end{equation}
\underline{without any imposed symmetry}. 
It may be assumed either left-handed or right-handed, because it is 
self-dual.  It may then connect to $(\nu,e)_L$ through 
$\Phi_R$, in which case no new automatic symmetry arises.  However, 
if $\xi_R$ is used instead, there is no such connection.  Hence 
the fermion bidoublet $\psi$ has its own conserved $U(1)$ symmetry, 
with $\psi^0_1 \psi^0_2 - \psi^-_1 \psi^+_2$ as an invariant 
mass term.  Again it can be shown~\cite{s95} that the neutral member 
$\psi^0_{1,2}$ is lighter than the charged member, so that the former 
is a simple predestined dark-matter candidate.  Now it appears that  
$\psi^0_{1,2}$ is a Dirac fermion and couples to the SM $Z$ boson. 
It would then be ruled out by direct-detection experiments by many orders of 
magnitude.  However, as already pointed out~\cite{gh16}, this $U(1)$ is 
broken radiatively to $Z_2$ from $W_L-W_R$ mixing.  Hence $\psi^0_{1,2}$ 
splits up to two Majorana fermions and the direct-detection limit from 
the $Z$ boson does not apply.

Consider next a fermion triplet under $SU(2)_L$ or $SU(2)_R$:
\begin{eqnarray}
\Sigma_L &=& (\Sigma_L^+,\Sigma_L^0,\Sigma_L^-) \sim (1,3,1,0), \\ 
\Sigma_R &=& (\Sigma_R^+,\Sigma_R^0,\Sigma_R^-) \sim (1,1,3,0). 
\end{eqnarray}
If $\Phi_R$ is available, then $\Sigma_R$ may be connected to $(\nu,e)_R$ 
and would not generate any new symmetry.  However, the absence~\cite{admnw09} 
of an $SU(2)_L$ scalar doublet $\Phi_L$ means that $\Sigma_L$ would have a 
conserved symmetry, i.e. $Z_2$ because $\Sigma_L$ is a Majorana triplet 
\underline{without any imposed symmetry}.  Note that $\Sigma^0_L$ does not 
couple to the SM $Z$ or the scalar bidoublet $\eta$.  Its only 
interaction is with $\Sigma_L^\pm$ through the SM $W_L^\pm$ bosons. 
This would allow it to become a viable dark-matter candidate as shown 
some years ago~\cite{ms09}.  The important difference is that $Z_2$ 
was imposed in that model of scotogenic neutrino mass, whereas here 
it is predestined. 

For the choice of an scalar triplet $\xi_R$ in breaking $SU(2)_R$, 
$\Sigma_R$ does not connect to $(\nu,e)_R$.   Hence $\Sigma_R^0$ is now 
also a viable dark-matter candidate.  Its presence has been 
discussed~\cite{hp15,ahr17}, which may also be motivated 
by $SO(10)$ unification. 
For completeness, a fermion singlet
\begin{equation}
S \sim (1,1,1,0) 
\end{equation}
would also have an automatic discrete $Z_2$ symmetry in this case. 
However, $S$ has no interaction by itself and would not have the correct 
dark-matter relic abundance from thermal freezeout.

As for possible scalar dark matter, consider again a triplet under 
$SU(2)_L$ or $SU(2)_R$:
\begin{eqnarray}
\chi_L &=& (\chi_L^+,\chi_L^0,\chi_L^-) \sim (1,3,1,0), \\ 
\chi_R &=& (\chi_R^+,\chi_R^0,\chi_R^-) \sim (1,1,3,0).
\end{eqnarray}
If $\Phi_R$ or $\xi_R$ is available, then $\chi_R$ couples to the triplet 
decomposition of $\Phi_R^\dagger \Phi_R$ or $\xi_R^\dagger \xi_R$ and would 
not generate any new symmetry.  Conversely, if $\Phi_L$ or $\xi_L$ is absent, 
$\chi_L$ would have a conserved $Z_2$ symmetry, and the real scalar 
$\chi_L^0$ would be predestined dark matter. 
 
The results of this section are summarized in Table 1.
\begin{table}[htb]
\caption{Predestined dark matter (PDM) from left-right scalar content.}
\begin{center}
\begin{tabular}{|c|c|c|c|c|c|}
\hline
$\eta$ & $\Phi_R$ & $\Phi_L$ & $\xi_R$ & $\xi_L$ & PDM \\ 
\hline
$\surd$ & $\surd$ & $\surd$ & -- & -- & -- \\ 
$\surd$ & $\surd$ & -- & -- & -- & $\Sigma_L, \chi_L$ \\ 
\hline
$\surd$ & -- & -- & $\surd$ & $\surd$ & $S,\psi,\Sigma_{L,R}$ \\ 
$\surd$ & -- & -- & $\surd$ & -- & $S,\psi,\Sigma_{L,R},\chi_L$ \\ 
\hline
\end{tabular}
\end{center}
\end{table}
Of the possible dark-matter candidates, the neutral fermion or scalar 
member of an $SU(2)_L$ triplet is the most viable because it interacts 
with its charged member through the SM $W_L^\pm$ boson.  Higher multiplets 
are of course also possible.  For example, an $SU(2)_R$ fermion quintet 
has recently been considered~\cite{agp18}.

\noindent \underline{\it Nonminimal Left-Right Extensions}~:~
The left-right model may be embedded in $SO(10)$ in which case there is no 
need for new fermions.  It may also be embedded in $E_6$ using its 
fundamental \underline{27} representation which decomposes into 
$(3,3^*,1)+(1,3,3^*)+(3^*,1,3)$ under its maximal subgroup  
$SU(3)_C \times SU(3)_L \times SU(3)_R$:
\begin{equation}
\pmatrix{d & u & h \cr d & u & h \cr d & u & h}, ~~~ 
\pmatrix{\psi_1^0 & \psi_2^+ & \nu \cr \psi_1^- & \psi_2^0 & e \cr 
\nu^c & e^c & S}, ~~~ 
\pmatrix{d^c & d^c & d^c \cr u^c & u^c & u^c \cr h^c & h^c & h^c}.
\end{equation}
Hence the particle content has automatically new fermions.  In terms of 
$SO(10) \times U(1)_\psi$, the \underline{27} decomposes to 
$(16,1)+(10,-2)+(1,4)$.  The further decomposition to 
$SU(5) \times U(1)_\chi$ is then 
$16 = (5^*,3)+(10,-1)+(1,-5)$, $10 = (5^*,-2)+(5,2)$ and $1 = (1,0)$.  
Consider the assignments of the fundamental fermions of Eq.~(10) 
under $U(1)_\psi$ and $U(1)_\chi$:
\begin{eqnarray}
U(1)_\psi &:& \pmatrix{1 & 1 & -2 \cr 1 & 1 & -2 \cr 1 & 1 & -2}, ~~~ 
\pmatrix{-2 & -2 & 1 \cr -2 & -2 & 1 \cr 1 & 1 & 4}, ~~~ 
\pmatrix{1 & 1 & 1 \cr 1 & 1 & 1 \cr -2 & -2 & -2}, \\ 
U(1)_\chi &:& \pmatrix{-1 & -1 & 2 \cr -1 & -1 & 2 \cr -1 & -1 & 2}, ~~~ 
\pmatrix{-2 & 2 & 3 \cr -2 & 2 & 3 \cr -5 & -1 & 0}, ~~~ 
\pmatrix{3 & 3 & 3 \cr -1 & -1 & -1 \cr -2 & -2 & -2}.
\end{eqnarray}
This shows that the known quarks and leptons are odd while the other fermions 
are even under $U(1)_\psi$.  Calling this charge $D_\psi$, the derived 
parity~\cite{m15}
\begin{equation}
R_\psi = (-1)^{D_\psi + 2j},
\end{equation}
where $j = 1/2$ is the spin of the particle, may then be used as the 
stabilizing dark parity of this model.  Note that the singlet fermion $S$ and 
the bidoublet fermion $\psi$ discussed earlier are already part of 
the $(1,3,3^*)$ multiplet.  Hence this is an example of possible predestined 
dark matter within a given gauge symmetry and its particle content.
However, it has been pointed out recently~\cite{kmppz18-2} that 
in the course of breaking $[SU(3)]^3$ to the SM using only bifundamental 
scalars, $R_\psi$ is only preserved with an imposed $Z_2$ symmetry.  
This invalidates $S$ or $\psi$ as predestined dark matter. Another 
closely related example is also known~\cite{dhqvv18}.

Under $U(1)_\chi$, there appears to be a possible residual $U(1)$ symmetry, 
under which $h,\psi^+_2,\psi^0_2$ and $h^c,\psi_1^0,\psi_1^-$ 
transform oppositely, whereas others are trivial.  Again the bifundamental 
scalars used in the symmetry breaking would spoil this scenario without 
an imposed symmetry.

Because there are two copies of $5^*$ of $SU(5)$ in $E_6$, an alternative 
left-right model (ALRM)~\cite{m87} is possible.  At the 
$SU(2)_L \times SU(2)_R \times U(1)_X$ 
level, the quarks and leptons are then given by
\begin{eqnarray}
&& \pmatrix{u \cr d}_L \sim (2,1,{1 \over 6}), ~~~ 
\pmatrix{u \cr h}_R \sim (1,2,{1 \over 6}), 
~~~ d_R \sim (1,1,-{1 \over 3}), ~~~ h_L \sim (1,1,-{1 \over 3}), \\ 
&& \pmatrix{\nu &\psi^+_2 \cr e & \psi^0_2}_L \sim (2,2,0), ~~~ 
\pmatrix{S \cr e}_R \sim (1,2,-{1 \over 2}), ~~~ \pmatrix{\psi_1^0 \cr 
\psi_1^-}_L \sim (2,1,-{1 \over 2}).
\end{eqnarray}
The would-be $\bar{h}_L d_R$ term and the coupling of the two lepton fermion 
doublets through the $\eta$ scalar bidoublet are forbidden by an assumed 
$Z_2$ or $U(1)$ symmetry, the origin of which in the case 
of $[SU(3)]^3$ was discussed already~\cite{kmppz18-2}.   This structure 
was recognized originally~\cite{m87} to eliminate the presence of flavor 
changing neutral currents, but was subsequently used to accommodate 
dark matter in a simpler version where $(\nu,e)$ is an $SU(2)_L$ 
doublet without being part of an $SU(2)_L \times SU(2)_R$ bidoublet. 
In these dark left-right models (DLRM)~\cite{klm09,klm10,bmw14}, $S$ is 
a possible dark-matter candidate but its stability depends on an imposed 
global $U(1)$ symmetry.  Recently, a model with an imposed gauge $U(1)$ 
symmetry has also been proposed~\cite{kmppz18-1}.

The DLRM particle content is naturally embedded in an $[SU(3)]^4$ 
model~\cite{bmw04} where leptonic color $SU(3)_l$~\cite{fl90,flv91} has 
been added.  Under $SU(3)_q \times SU(3)_L \times SU(3)_l \times SU(3)_R$, 
the fermion chain is given by
\begin{equation}
\pmatrix{d & u & h \cr d & u & h \cr d & u & h}, ~~~ 
\pmatrix{x_1 & x_2 & \nu \cr y_1 & y_2 & e \cr z_1 & z_2 & n}, ~~~ 
\pmatrix{z_1^c & y_1^c & x_1^c \cr z_2^c & y_2^c & x_2^c \cr 
n^c & e^c & \nu^c}, ~~~ 
\pmatrix{h^c & h^c & h^c \cr u^c & u^c & u^c \cr d^c & d^c & d^c},
\end{equation}
where the $SU(2)_R$ fermion doublet is now denoted $(e^c,n^c)$, with 
$n^c$ a dark-matter candidate, again after an imposed symmetry.
Here $SU(3)_l$ has an unbroken $SU(2)_l$ subgroup which serves to confine 
the half-charged hemions $(x,y,z)$ in the same way that $SU(3)_q$ confines 
the quarks with one-third and two-third charges.  Their phenomenology at 
a future $e^-e^+$ collider has been discussed
recently~\cite {kmppz17,kmppz18-3}.  It was also ascertained that the 
scalar singlet in an $(1,3,1,3^*)$ multiplet is a better dark-matter 
candidate.

There is another $[SU(3)]^4$ version~\cite{m05} without leptonic color. 
The extra $SU(3)_D$ allows two fermion bidoublets, i.e. 
$SU(2)_L \times SU(2)_D$ and $SU(2)_R \times SU(2)_D$, which 
have the same form as $\psi$ with neutral components as  
dark matter.  The former is ruled out because the neutral components 
couple to the SM Z boson, but the latter could work.  This idea has only 
been implemented recently in an explicit 
$SU(3)_C \times U(1)_{B-L} \times SU(2)_L \times SU(2)_R \times SU(2)_D$ 
model~\cite{m18}.  It is a genuine first example of predestined $U(1)$ dark 
matter, in analogy to baryon and lepton numbers but unrelated to them.

\noindent \underline{\it Gauge $U(1)$ Extensions}~:~
The best known gauge $U(1)$ extension is just $SU(3)_C \times SU(2)_L \times 
U(1)_Y \times U(1)_{B-L}$, under which 
\begin{eqnarray}
&& \pmatrix{u \cr d}_L \sim (3,2,{1 \over 6},{1 \over 3}), ~~~ 
u_R \sim (3,1,{2 \over 3},{1 \over 3}), ~~~ 
d_R \sim (3,1,-{1 \over 3},{1 \over 3}), \\ 
&& \pmatrix{\nu \cr e}_L \sim (1,2,-{1 \over 2},-1), ~~~ 
e_R \sim (1,1,-1,-1), ~~~ \nu_R \sim (1,1,0,-1).
\end{eqnarray}
Note that $\nu_R$ is required for $U(1)_{B-L}$ to be anomaly-free. 
To break $U(1)_{B-L}$, a neutral singlet $\zeta$ with 2 units of $B-L$ charge 
is the conventional choice, in which case $\nu_R$ gets a large Majorana 
mass, and the canonical seesaw mechanism allows $\nu_L$ to acquire a 
small Majorana mass.  If $\zeta$ has 3 units of $B-L$ 
charge~\cite{mpr13,ms15,mpsz15}, then $\nu$ is a Dirac neutrino.
In either case, $B$ is conserved.  The one Higgs doublet 
$\Phi \sim (1,2,1/2,0)$ is the same as in the SM.

Consider now a fermion singlet or triplet which is also trivial under 
$U(1)_{B-L}$:
\begin{equation}
S \sim (1,1,0,0), ~~~ \Sigma = (\Sigma^+, \Sigma^0, \Sigma^-) \sim 
(1,3,0,0).
\end{equation}
Neither would couple to $\Phi$ and a lepton doublet because of $U(1)_{B-L}$. 
Hence they are predestined dark matter.  Again $S$ is completely decoupled, 
but $\Sigma$ works well as dark matter, as already mentioned in the case 
of left-right models without an $SU(2)_L$ scalar doublet. 

As for scalar dark matter, assuming that $\zeta \sim (1,1,0,n)$ is used 
to break $U(1)_{B-L}$, then a scalar triplet
\begin{equation}
\chi = (\chi^+,\chi^0,\chi^-) \sim (1,3,0,n')
\end{equation}
would have an $U(1)$ dark symmetry if $n'$ is not zero, or $\pm n$.  
In that case, $\chi^0$ would couple to the $U(1)_{B-L}$ gauge boson as well 
as the SM Higgs boson, and be seriously constrained by present data.
If a scalar singlet dark-matter candidate is desired, then it should be 
charged under $U(1)_{B-L}$, but not $\pm n$ or $\pm 2n$ or 
$\pm 3n$~\cite{ry15}.

Instead of $U(1)_{B-L}$ to support Majorana seesaw neutrino masses with 
heavy right-handed singlets $\nu_R$, a peculiar anomaly-free $U(1)_X$ 
may be used with heavy triplet fermions
\begin{equation}
\Sigma_R = (\Sigma^+, \Sigma^0, \Sigma^-)_R \sim (1,3,0,n_6).
\end{equation}
Together with 
\begin{eqnarray}
&& \pmatrix{u \cr d}_L \sim (3,2,{1 \over 6},n_1), ~~~ 
u_R \sim (3,1,{2 \over 3},n_2), ~~~ 
d_R \sim (3,1,-{1 \over 3},n_3), \\ 
&& \pmatrix{\nu \cr e}_L \sim (1,2,-{1 \over 2},n_4), ~~~ 
e_R \sim (1,1,-1,n_5),
\end{eqnarray}
it was shown some years ago~\cite{m02,mr02,bd05} and studied more 
recently~\cite{aem09,abm16} that for $3n_1 + n_4 \neq 0$, an anomaly-free 
solution exists per family if 
\begin{equation}
n_2 = {1 \over 4} (7n_1-3n_4), ~~~ n_3 = {1 \over 4}(n_1+3n_4), ~~~ 
n_5 = {1 \over 4} (-9n_1+5n_4), ~~~ n_6 = {1 \over 4} (3n_1+n_4).
\end{equation}
The scalar sector consists of a singlet $(1,1,0,(3n_1+n_4)/2)$ and 
two doublets $(1,2,1/2,3(n_1-n_4)/4)$ and $(1,2,1/2,(9n_1-n_4)/4)$. 
The singlet and triplet fermions of Eq.~(19) are thus also predestined 
dark matter because neither would couple to either scalar doublet and a 
lepton doublet unless $3n_1+n_4=0$.  As for the scalar triplet of Eq.~(20), 
an $U(1)$ dark symmetry would emerge if $n'$ is not zero, or 
$\pm(3n_1+n_4)/2$, or $\pm(3n_1+n_4)$.  Again, $\chi^0$ would then couple 
to the $U(1)_X$ gauge boson as well as the two Higgs doublets, and be 
seriously constrained by present data.

\noindent \underline{\it Concluding Remarks}~:~
In well-motivated gauge extensions of the SM, there are often simple fermion 
or scalar multiplets, i.e. singlets, bidoublets, and triplets, which may 
be stable \underline{without any imposed symmetry}.  Their neutral 
components are often automatically the lightest and thus predestined 
to be dark matter.  In this paper, several examples are discussed in the 
context of left-right and $U(1)$ extensions of the SM.  In the former, 
many examples are already known~\cite{gh16,hp15,ahr17,agp18} which involve 
mostly fermions.  It is pointed out here that the absence of an $SU(2)_L$ 
scalar doublet~\cite{admnw09} allows the $SU(2)_L$ scalar triplet $\chi_L$ 
of Eq.~(8) to be dark matter as well.  In the latter, whether $U(1)_{B-L}$ 
or $U(1)_X$ is used, there are also fermion and scalar triplets which could 
be predestined dark matter.  The stabilizing dark symmetry is either $Z_2$ 
or $U(1)$.

\noindent \underline{\it Acknowledgement}~:~
This work was supported in part by the U.~S.~Department of Energy Grant 
No. DE-SC0008541.

\baselineskip 18pt
\bibliographystyle{unsrt}

\end{document}